\documentclass[symmetry,article,submit,moreauthors,pdftex]{mdpi}
\usepackage{bbm}
\usepackage{bm}
\usepackage{physics}
\usepackage{amssymb}
\usepackage{scalerel}
\usepackage{datetime} 
\usepackage{siunitx}
\usepackage{comment}

\firstpage{1}
\pubvolume{1}
\issuenum{1}
\articlenumber{0}
\pubyear{2021}
\copyrightyear{2021}
\datereceived{}
\dateaccepted{}
\datepublished{}
\hreflink{https://doi.org/
} 
\Title{A search for neutron to mirror neutron oscillation using neutron electric dipole moment measurements}

\Author{
Prajwal Mohanmurthy $^{1,2,\orcidA{},}$*,
Albert R. Young $^{3,}$*
Jeff A. Winger $^{4,}$,
Geza Zsigmond $^{5,}$
}

\address{%
$^{1}$ \quad Department of Physics, University of Chicago, Chicago, IL 60637, U.S.A\\
$^{2}$ \quad Laboratory for Nuclear Science, Massachusetts Institute of Technology, MA 02139, U.S.A\\
$^{3}$ \quad Department of Physics, North Carolina State University, Raleigh, NC 27695, U.S.A\\
$^{4}$ \quad Department of Physics and Astronomy, Mississippi State University, Mississippi State, MS 39762, U.S.A\\
$^{5}$ \quad Laboratory for Particle Physics, Paul Scherrer Institute, PSI-Villigen, 5233, Switzerland
}

\corres{E-mail: prajwal@mohanmurthy.com, aryoung@ncsu.edu}

\abstract{
Baryon number violation is a key ingredient of baryogenesis. It has been hypothesized that there could also be a parity-conjugated copy of the standard model particles, called mirror particles. The existence of such a mirror universe has specific testable implications, especially in the domain of neutral particle oscillation, \emph{viz.} the baryon number violating neutron to mirror-neutron ($n-n'$) oscillation. Consequently, there were many experiments that have searched for $n-n'$ oscillation, and imposed constraints upon the parameters that describe it. Recently, further analysis on some of these results have identified anomalies which could point to the detection of $n-n'$ oscillation. All the previous efforts searched for $n-n'$ oscillation by comparing the relative number of ultracold neutrons that survive after a period of storage for one or both of the two cases: (i) comparison of zero applied magnetic field to a non-zero applied magnetic field, and (ii) comparison where the orientation of the applied magnetic field was reversed. However, $n-n'$ oscillations also lead to variations in the precession frequency of polarized neutrons upon flipping the direction of the applied magnetic field. Precession frequencies are measured, very precisely, by experiments searching for the electric dipole moment. For the first time, we used the data from the latest search for the neutron electric dipole moment [Phys. Rev. Lett. 124, 081803 (2020)] to constrain $n-n'$ oscillation. After compensating for the systematic effects that affect the ratio of precession frequencies of ultracold neutrons and cohabiting $^{199}$Hg-atoms, chief among which was due to their motion in non-uniform magnetic field, we constrained any further perturbations due to $n-n'$ oscillation. We thereby provide a lower limit on the $n-n'$ oscillation time constant of $\tau_{nn'}/\sqrt{|\cos(\beta)|} > \SI{5.7}{~s},~\SI{0.36}{~\micro T'}<B'<\SI{1.01}{~\micro T'}$ (95\% C.L.), where $\beta$ is the angle between the applied magnetic field and the ambient mirror magnetic field. This constraint is the best available in the range of $\SI{0.36}{~\micro T'}<B'<\SI{0.40}{~\micro T'}$.
}



\begin{document}


\section{Introduction}
It had already been noted by Lee and Yang, in their landmark paper \cite{Lee1956-ai}, that observation of apparent parity violation could be suppressed with the introduction of a parity-conjugated copy of the same weakly interacting particles. It was later shown by Kobzarev, Okun, and Pomeranchuk \cite{Kobzarev1966-bt} that normal matter would not interact via the strong, weak, or electromagnetic interactions with their \emph{mirror} counterparts. Nonetheless, mirror particles could have their own set of interactions separate from the electromagnetic, weak, and strong interactions of the Standard Model (SM). However, there may be interactions beyond the SM (BSM), between neutral SM particles and their mirror counterparts (SM$^{\prime}$), which mixes the neutral SM particles with their SM$^{\prime}$ counterparts. Foot and Volkas showed that with the introduction of mirror matter, parity and time reversal symmetries could be restored in the global sense \cite{Pavsic1974-hz,Foot1991-xw,Foot1992-ko}.

Mixing of SM and SM$^{\prime}$ particles can provide answers for several standing issues in physics today. Mirror matter can act as dark matter candidate \cite{Khlopov1991-xx,Hodges1993-fx,Foot2003-xq,Foot2004-pn,Berezhiani2005-ic,Berezhiani2006-gd}. Interpretations, involving mirror matter, of the anomalies in experiments searching for dark matter can be found in Refs.~\cite{Foot2010-ro,Foot2012-nn,Addazi2015-xw,Cerulli2017-ea}. Gauge bosons \cite{Berezhiani1998-qw} can also induce mixing between neutral mesons that could be used to search for dark matter \cite{Berezhiani2008-gd}. Oscillation between neutrinos and mirror-neutrinos could provide a candidate sterile neutrino in the form of mirror-neutrinos \cite{Zeldovich1981-ii,Akhmedov1992-px,Berezhiani1995-td,Foot1995-xz,Silagadze1997-po,Berezinsky2003-rp}. Similarly, mixing of photons and mirror-photons has also been proposed \cite{Holdom1986-ha,Glashow1986-fy,Foot2001-ee,Berezhiani2009-uc}, and tested via positronium to mirror-positronium oscillation \cite{Gninenko1994-nq,Foot2000-ku,Gninenko2004-rp,Gninenko2006-un,Badertscher2007-hs,Crivelli2010-jk,Vigo2018-ut}.

Baryogenesis requires baryon number violation, CP violation, and interaction out of the thermal equilibrium \cite{Sakharov1967-vt,Kuzmin1970-kx}. Mixing of SM and SM$^{\prime}$ particles also could provide additional channels for CP violation, thereby contributing to baryo/leptogenesis and the baryon asymmetry of the universe \cite{Nishijima1965-ue,Berezhiani2001-ey,Bento2001-ol,Bento2002-em}. Neutron to mirror-neutron oscillation is one such process \cite{Berezhiani2016-pw}. Neutron to mirror-neutron oscillations could provide a mechanism to relax the Greisen-Zatsepin-Kuzmin (GZK) limit \cite{Berezhiani2012-us,Berezhiani2006-zn}, have implications upon neutrons from solar flares \cite{Mohapatra2005-ux}, and affect neutron stars \cite{Berezhiani2016-dz,Berezhiani2021-xh}. Neutron to mirror-neutron oscillations have also been discussed in the context of neutron to anti-neutron oscillation \cite{Mohapatra1980-jy,Baldo-Ceolin1994-ih,Phillips2016-aq,Berezhiani2015-ah,Berezhiani2016-mb,Addazi2021-vk}. 

The existence of mirror matter has cosmological implications upon the formation of the universe, including through baryogenesis and nucleosynthesis \cite{Blinnikov1982-wn,Kolb1985-pv,Carlson1987-qp,
Ciarcelluti2005-px,Ciarcelluti2005-fd,Das2011-qb,Coc2013-rp,Coc2014-mb}. Mirror matter itself could also evolve to form large scale structures \cite{Mohapatra1997-li,Foot1999-ha,Foot1999-gm,Ignatiev2003-na,Foot2004-yt,Berezhiani2010-gv}. Various other cosmological implications of mirror matter and mixing between the SM and SM$^{\prime}$ particles have also been well explored \cite{
Berezhiani1996-kf,Berezhiani2004-ui,Das2011-rv,Dvali2009-dv,Foot2014-py}. Comprehensive reviews involving mirror matter can be found in Refs.~\cite{Berezhiani2005-lq,Okun2007-ge,Berezhiani2008-lv,Dubbers2011-kn,Berezhiani2018-bi} and in the references therein.



Berezhiani and Bento pointed out that the characteristic time associated with neutron to mirror-neutron oscillations ($n-n'$) can be on the order of few seconds, and ergo small compared to the lifetime of the neutron \cite{Berezhiani2006-sj}. The theory of $n-n'$ oscillation was further detailed in Ref.~\cite{Berezhiani2009-hb}, in which Berezhiani showed that, as long as neutrons and their mirror counterparts have the same mass and decay widths, application of a magnetic field equal to the ambient mirror magnetic field can induce a degeneracy between the $\ket{n}$ and $\ket{n'}$ states, resonantly enhancing the $n-n'$ oscillation. The $n-n'$ oscillation can be described by a $4\times4$ Hamiltonian \cite{Berezhiani2009-hb},
\begin{eqnarray}
\mathcal{H} = \begin{bmatrix}
\mu_n \bm{B \cdot \sigma} & 1/\tau_{nn'} \\
1/\tau_{nn'} & \mu_n \bm{B' \cdot \sigma}
\end{bmatrix},\label{eq10-1}
\end{eqnarray}
where $\mu_n=-60.3~$neV/T is the magnetic moment of the neutron, $\tau_{nn'}$ is the time constant for the $n-n'$ oscillation, ${\bm{B^{(')}}}$ is the (mirror) magnetic field, and $\bm{\sigma} = \langle \sigma_x, \sigma_y, \sigma_z\rangle$, the $2\times2$ Pauli matrices with $\hbar=1$. 

Neutron to mirror-neutron oscillation affects the precession frequency of neutrons under the influence of a magnetic field. The probability that a neutron flips its polarization state from aligned to anti-aligned, \emph{w.r.t.} the magnetic field, after time $t$, for the Hamiltonian in Eq.~\ref{eq10-1}, is given by \cite{Berezhiani2009-hb}
\begin{eqnarray}
P^{\scaleto{(+/-)}{4pt}}_{BB'}(t)&=&
~~\cos^4\left(\theta\right)\sin^2\left(2\phi\right)\sin^2\left(2\tilde{\omega}t\right) \nonumber \\
&&+\frac{1}{2}\sin^2\left(2\theta\right)\sin\left(2\phi\right)\sin\left(2\phi'\right)\sin\left(\tilde{\omega}t\right)\sin\left(\tilde{\omega}'t\right) \nonumber \\
&&+\sin^4\left(\theta\right)\sin^2\left(2\phi'\right)\sin^2\left(2\tilde{\omega}'t\right)
,~\label{eq10-3}
\end{eqnarray}
where the angles, $\theta$, $\phi$, and $\phi'$ are defined by
\begin{eqnarray}
\tan\left(2\theta\right)&=&\frac{1}{\tau_{nn'}}\frac{|\bm{\omega}+\bm{\omega'}|}{\omega'^2-\omega^2}=\frac{1}{\tau_{nn'}\omega'}\sqrt{\frac{1+\eta^2+2\eta\cos(\beta)}{(1-\eta^2)^2}},~\label{eq10-4} \\
\sin\left(2\phi\right)&=&\frac{|\bm{\omega}\times\bm{\omega'}|}{\tilde{\omega}|\bm{\omega}+\bm{\omega'}|},~\label{eq10-5} \\
\sin\left(2\phi'\right)&=&\frac{|\bm{\omega}\times\bm{\omega'}|}{\tilde{\omega}'|\bm{\omega}+\bm{\omega'}|}\quad,~\label{eq10-6}
\end{eqnarray}
and, $\eta=B/B'$, $\bm{\omega^{(')}} = \mu_n\bm{B^{(')}}/2 = \SI{45.81}{~(\micro T}\cdot\text{s})^{-1}B^{(')}$, and $\beta$ is the angle between $\bm{B}$ and $\bm{B'}$. Clearly from Eq.~\ref{eq10-4} it can be seen that the angle $\theta$ sets the magnitude of the $n-n'$ oscillation. The angles of $\phi$ and $\phi'$ describe the spin precession and are eliminated when the probabilities are time averaged. In Eqs.~\ref{eq10-4}-\ref{eq10-6}, $\tilde{\omega}$ and $\tilde{\omega}'$ are the eigenvalues of the Hamiltonian in Eq.~\ref{eq10-1}, given by
\begin{eqnarray}
\tilde{\omega}&=&\frac{1}{2}\sqrt{\frac{1}{\tau^2_{nn'}}+2\left(\omega^2+\omega'^2\right)+2\left(\omega^2-\omega'^2\right)\sqrt{1+\tan^2\left(2\theta\right)}},~\label{eq10-7} \\
\tilde{\omega}'&=&\frac{1}{2}\sqrt{\frac{1}{\tau^2_{nn'}}+2\left(\omega^2+\omega'^2\right)-2\left(\omega^2-\omega'^2\right)\sqrt{1+\tan^2\left(2\theta\right)}}\quad.~\label{eq10-8}
\end{eqnarray}

When there is no $n-n'$ oscillation, the expression in Eq.~\ref{eq10-3} reduces to show the usual magnetic field dependent Larmor spin precession, $P^{\scaleto{(+/-)}{4pt}}_{BB'}(t)=\sin^2\left(2\phi\right)\sin^2\left(2\omega t\right)$. But in the presence of $n-n'$ oscillation, the neutron may precess, even when the magnetic field is zero. Also, in the presence of $n-n'$ oscillation, the precession frequency varies upon reversing the direction of the magnetic field. For the case where, $\theta\ll1$, and $(\bm{B}-\bm{B'})\gg0$, \emph{i.e.} far away from resonance, this difference in precession frequency for the neutron-like state is given by \cite{Berezhiani2009-hb}
\begin{eqnarray}
\frac{\delta\omega}{\omega} = \frac{\cos\left(\beta \right)}{\tau_{nn'}^2}f_{d}(\eta)~;~f_{d}(\eta) = \frac{1}{2\omega'^2\eta\left(\eta^2-1\right)},~\label{eq10-9}
\end{eqnarray}
where $f_{d}(\eta)$ is a scaling function dependent on the applied magnetic field and the ambient mirror magnetic field. Eq.~\ref{eq10-9} has a singularity around $B'=B$, and the sensitivity of the experiment drops when using a magnetic field that is farther away from the ambient mirror magnetic field. Near resonance, where $(\bm{B}-\bm{B'})\sim0$, an analytic approximation is used to time average the square of sinusoidal function in Eq.~\ref{eq10-3} \cite{Ignatiev2000-ah,Berezhiani2018-bi} to obtain
\begin{eqnarray}
\langle\sin^2\left(\omega t\right)\rangle\approx\frac{1}{2}\left( 1- \exp\left(-2\omega^2\sigma^2_f\right)\cos\left(2\omega\langle t_f \rangle\right)\right),~\label{eq10-9b}
\end{eqnarray}
where $\sigma_f = \langle t^2_f \rangle - \langle t_f \rangle^2$, and $\langle t_f \rangle$ is the mean free-flight time between two consecutive wall collisions. However, a more detailed modeling of precession signals near the resonance field region needs to be investigated.


The local mirror magnetic field at the site of the experiment, may be nonzero \cite{Berezhiani2009-hb,Berezhiani2012-rq}. A local mirror magnetic field may have terrestrial \cite{Ignatiev2000-ah,Berezhiani2012-rq} or galactic origins \cite{Foot2004-yt,Berezhiani2004-wx}. In the case of terrestrial origins, such mirror magnetic fields may be bound to the reference frame of the Earth \cite{Berezhiani2004-ui}, in which case the angle $\beta$ is constant. On the other hand, such mirror magnetic fields may have galactic origins, in which case the angle $\beta$ varies sidereally \cite{Berezhiani2009-hb}. 
In this work, we focus on the case where the mirror magnetic field is assumed to be fixed to the reference frame of the Earth.

The neutron has a large magnetic moment, and a non-zero electric dipole moment \cite{Khriplovich2012-no,Pospelov2005-xn}. In order to measure the electric dipole moment (EDM) of the neutron, most modern experiments have employed ultracold neutrons (UCNs), while using the Ramsey technique of separated oscillating fields \cite{Ramsey1950-ma} at room temperature. The magnetic moment, as well as the electric dipole moment, affect the precession frequency of the neutrons subject to both magnetic and electric fields simultaneously, as seen in the equation 
\begin{eqnarray}
\hbar\omega^{B_{\uparrow},E_{U/D}} &=& 2|\bm{\mu}\cdot\bm{B} \pm \bm{d}\cdot\bm{E}|,~\label{eq10-10}
\end{eqnarray}
where the superscript indices denote the magnetic or electric field along with their relative direction. Here, the magnetic field is held constant, while the flipping of the direction of the electric field leads to the variation of the precession frequency that is linked to the electric dipole moment. The shifts in precession frequency upon reversing magnetic field direction, on the other hand, can also be measured in neutron EDM experiments, and is sensitive to $n-n'$ oscillation according to Eq.~\ref{eq10-9}. In this work, we have used the neutron precession frequency shifts upon flipping the direction of the magnetic field as a means to gain sensitivity to $n-n'$ oscillation in experiments measuring the neutron EDM.

\section{Prior measurements in search of $n-n'$ oscillation}


Mixing of a neutron and its mirror counterpart allows for the possibility to directly search for $n-n'$ oscillation in experiments \cite{Pokotilovski2006-mx,Kerbikov2008-uk}. All but one of the experiments in search of $n-n'$ oscillation have employed the disappearance technique ($n\rightarrow n'$). Under the assumption of $B'=0$, the first experiments set the constraints of $\tau_{nn'} > 103~$s (95\% C.L.) \cite{Ban2007-dt} and $\tau_{nn'} > 414~$s (90\% C.L.) \cite{Serebrov2008-jg}, which has since been updated to $\tau_{nn'} > 448~$s (90\% C.L.) \cite{Serebrov2009-yf}. Refs.~\cite{Altarev2009-bv,Bodek2009-py} relaxed the condition of $\bm{B'}=0$, and set a constraint of $\tau_{nn'} > 12~\text{s},~\SI{0.4}{~\micro T'}<B'<\SI{12.5}{~\micro T'}$ (95\% C.L.). However, Refs.~\cite{Berezhiani2009-hb,Berezhiani2012-rq} identified that a reanalysis of the above experiments resulted in two statistically significant anomalies indicating $n-n'$ oscillation. A recent update in Refs.~\cite{Berezhiani2018-df,Biondi2018-cu} shows a persistence of these anomalies. Ref.~\cite{Berezhiani2018-df} set a constraints of $\tau_{nn'}>17~\text{s},~\SI{8}{~\micro T'}<B'<\SI{17}{~\micro T'}$ (95\% C.L.) and $(\tau_{nn'}/\sqrt{\cos(\beta)})>27~\text{s}, \SI{6}{~\micro T'}<B'<\SI{25}{~\micro T'}$ (95\% C.L.). In addition to the two previously identified anomalies, Ref.~\cite{Berezhiani2018-df} also identified a third anomaly. The most recent dedicated effort, to test these anomalies, was conducted at PSI using the nEDM apparatus \cite{Abel2019-rk,Mohanmurthy2019-ju,Abel2021-hp}, and set the constraints of $\tau_{nn'}>352~\text{s}~\text{(95\% C.L.)}$, where $B'=0$, $\tau_{nn'}>6~\text{s},~\SI{0.36}{~\micro T'}<B'<\SI{25.66}{~\micro T'}$ (at 95\% C.L.), and $(\tau_{nn'}/\sqrt{\cos(\beta)})>9~\text{s},~\SI{5.04}{~\micro T'}<B'<\SI{25.39}{~\micro T'}$ (at 95\% C.L.).

An experiment using the reappearance technique ($n\rightarrow n' \rightarrow n$) has been demonstrated, and it set a constraint of $\tau_{nn'}>2.7~\text{s}~\text{(90\% C.L.)}$, where $B'=0$ \cite{Schmidt2007-dn}. Furthermore, there are proposals to improve both the disappearance \cite{Ayres2021-la}, and reappearance \cite{Broussard2017-ev,Broussard2019-gh} type measurements. In this work, we, for the first time, employ the method of searching for $n-n'$ oscillation by measuring the variations in precession frequency by flipping the direction of the magnetic field, using measurements from an experiment that searched for the neutron EDM.

\section{Measuring the precession frequency of the neutron}

A precession frequency shift which may be induced by $n-n'$ oscillation, when the magnetic field is flipped, would be detectable in experiments measuring the neutron EDM (nEDM). But the precession frequency of the neutron must be measured by applying both the directions of magnetic field. Furthermore, the magnetic field used must be close to the relevant mirror magnetic fields indicated in Refs.~\cite{Berezhiani2012-rq,Berezhiani2018-df}, \emph{i.e.} $B' \in (0.1,100)\SI{}{~\micro T'}$, so that such a measurement is sensitive to potential ambient mirror magnetic fields of interest. In the series of nEDM experiments culminating in Ref.~\cite{Serebrov2015-mg}, the magnetic field was fixed, while the series of nEDM experiments culminating in Refs.~\cite{Apostolescu1969-kv,Cohen1969-qw,Dress1977-yk} were performed using magnetic fields of $B_0=\{400,150,1700\}\SI{}{~\micro T}$, respectively. 

The most recent measurement of the neutron EDM was carried out at the Paul Scherrer Institute (PSI) in Switzerland. PSI nEDM collaboration performed its measurements by storing UCNs in a cylindrical chamber of height, $h=12.000(1)~$cm and radius, $R=23.500(1)~$cm, and by subjecting the UCNs to fields of $\langle B_0 \rangle\approx\pm\SI{1.036}{~\micro T}$ and $\langle E \rangle=\pm11~$kV/cm. In order to gain an extra handle on systematics, the PSI nEDM experiment measured the precession frequency of the neutrons by applying both directions for the magnetic field, with various magnetic field gradients. The mean magnetic field used is also in a region of interest indicated by Refs.~\cite{Berezhiani2012-rq,Berezhiani2018-df}. Further details about the apparatus \cite{Baker2014-hd,Afach2014-ji,Afach2015-lz,Ban2016-xw,Abel2020-ig}, data collection \cite{Abel2019-cv}, and characterization \cite{Abel2019-oh} have also been published. For the purposes of this work, we have used the measurements published in Ref.~\cite{Abel2020-jr}.


\subsection{Using co-magnetometer: measuring ratio of precession frequencies}

Measurement of the nEDM requires the magnetic field to be held constant, or to be measured very precisely and independently so that compensation can be made for any variations in the field strength. In order to account for drifts in the magnetic field, recent efforts have used a $^{199}$Hg vapor co-magnetometer that occupies the same volume as the stored UCNs to determine the ratio of precession frequencies, $\mathcal{R}=f_n/f_{\text{Hg}}$. Here, the neutron precession frequency ($f_n$) was measured independently from the precession frequency of the $^{199}$Hg atoms ($f_{\text{Hg}}$). The polarized neutron precession frequency was obtained from fitting the Ramsey resonance to the asymmetry between the number of spin-up and spin-down neutrons counted, while the precession frequency of the polarized $^{199}$Hg atoms was obtained from the intensity modulation of the transmitted light. The ratio of precession frequencies, $\mathcal{R}$, is affected by various sources (elaborated in the next sub-section) indicated by
\begin{eqnarray}
\mathcal{R}=\frac{\gamma_{n}}{\gamma_{\text{Hg}}}\left(1+\delta_{nn'}+\delta_{\text{grav}}+\left(\delta_{\text{T}}+\delta_{\text{Earth}}\right)+\delta^{\text{false}}_{\text{EDM}}+\mathcal{O}\right),~\label{eq10-11}
\end{eqnarray}
where $\gamma_i$ are the gyromagnetic ratios for the respective species in the subscript, and $\delta_{nn'}=\delta\omega/\omega$ from Eq.~\ref{eq10-9}. Neutron to mirror-neutron oscillation only affects free neutrons, with the relevant $\tau_{nn'}$ much greater than the collision time of the $n-n'$ superposition with the walls. The resultant shift in precession frequency, $\delta_{nn'}$ due to $n-n'$  oscillations should be averaged over the fields experienced by the $n-n'$ superposition. For this study, we take that average to be same as $B_0$. We expect a more detailed treatment would reflect the gravitational shift in the fields experienced by the mirror neutrons relative to the $^{199}$Hg co-magnetometer as well, but we have neglected it here. The precession frequency of $^{199}$Hg atoms effectively remains unaffected, since $n-n'$ oscillation is heavily suppressed inside a nuclei \cite{Addazi2017-hu}. Therefore, the ratio $\mathcal{R}$, preserves the influence of $n-n'$ upon the precession frequency of the neutrons, $f_n$, while also compensating for magnetic field drifts.

\subsection{Systematic corrections to $\mathcal{R}=f_n/f_{\text{Hg}}$}

The largest shift, $\delta_{\text{grav}}$, comes from a combination of the magnetic field gradient, $G_{\text{grav}}$, and the center-of-mass offset of the ensemble of UCNs \emph{w.r.t} the $^{199}$Hg vapor cloud, $\langle z \rangle$. The gravitational shift is given by \cite{Abel2019-oh,Abel2020-jr}
\begin{eqnarray}
\delta_{\text{grav}}=\frac{G_{\text{grav}}\cdot\langle z \rangle}{B_0}.~\label{eq10-11a}
\end{eqnarray}
Also, the analysis in Ref.~\cite{Abel2020-jr} accounted for drifts in the value of $G_{\text{grav}}$ as the data was collected.

The rotation of the Earth induces precession in the neutrons when measured in the non-inertial reference frame on the Earth \cite{Lamoreaux2007-ct}. The amount of precession frequency shift depends on the magnetic field applied and the latitude of the location on the Earth, but for a magnetic field of magnitude $\langle B_0 \rangle \approx\SI{1.036}{~\micro T}$ and at the latitude of PSI, $47.53648^{\circ}~$N, it is $\delta^{\uparrow/\downarrow}_{\text{Earth}}=\mp1.4\times10^{-6}$ \cite{Afach2014-ak}, where the arrows indicate the direction of the applied magnetic field. Residual transverse fields, $\bm{B_T}$, perpendicular to the axis of the UCN precession chamber, also cause a shift in the precession frequency given by $\delta_{\text{T}}=\langle B^2_T\rangle/(2B^2_0)$ \cite{Abel2019-oh}. Ref.~\cite{Abel2020-jr} uses a separate offline map of the magnetic field environment to determine the transverse fields. Both these shifts, $\delta_{\text{Earth}}$ and $\delta_{\text{T}}$, were accounted for by using a corrected value of the ratio of precession frequencies, $\mathcal{R}^{\text{corr}}=\mathcal{R}/(1+\delta_{\text{Earth}}+\delta_{\text{T}})$. The uncertainty in the precision of the measured values of $B_T$ translates to a dilution in sensitivity to the neutron EDM given in Table~\ref{tab10-1}.

False EDM effects, $\delta^{\text{false}}_{\text{EDM}}$, come mainly from linear shifts in the electric field due to relativistic motion in the field \cite{Pendlebury2004-yd} and magnetic field gradients, yielding 
\begin{eqnarray}
\delta^{\text{false}}_{\text{EDM}}=\frac{2E}{\hbar B_0 \gamma_n}\left(d^{\text{net}}_n+\frac{\hbar |\gamma_n \gamma_{\text{Hg}}|R^2}{8c^2}\left(G_{\text{grav}}+\hat{G}\right)\right),~\label{eq10-12}
\end{eqnarray}
where $\gamma_n/(2\pi)=29.164705(55)~$MHz/T \cite{Afach2014-ak} and $\gamma_{\text{Hg}}/(2\pi)=7.5901152(62)~$MHz/T \cite{Afach2014-ak}. False EDM effects due to net rotational motion of the UCNs inside the precession chamber was limited to $d^{\text{net}}_n<2\times10^{-28}~$e$\cdot$cm \cite{Abel2020-jr}. In Eq.~\ref{eq10-12}, $\hat{G}$ is higher order gradient terms which were measured for every configuration and subtracted by correcting the measured EDM, $d_n^{\text{corr}}=d_n-\hbar|\gamma_n \gamma_{\text{Hg}}|R^2\hat{G}/(8c^2)$, where $d_n$ is obtained using the measured precession frequencies of the neutrons. The corrections due to the higher order gradient term, were reported to move the values of measured neutron EDM by $(69\pm10)\times10^{-28}~$e$\cdot$cm \cite{Abel2020-jr}. Lastly, the EDM of $^{199}$Hg atoms, $d_{^{199}\text{Hg}}=(-2.20\pm2.75_{\text{stat}}\pm1.48_{\text{sys}})\times10^{-30}~$e$\cdot$cm \cite{Graner2016-ge}, translates to an uncertainty in the neutron EDM, and also shifts the value of $\mathcal{R}$ by an amount $\delta_{n\leftarrow\text{Hg}}=-2Ed_{^{199}\text{Hg}}/\hbar|\gamma_n|B_0$. A critical assumption in our analysis is that the dominant false EDM effect arising from the mercury co-magnetometer is insensitive to the couplings to the mirror sector.

Other effects in Eq.~\ref{eq10-11} mainly dilute the precision of the measurement of the neutron EDM, and also affect the uncertainty on the ratio of measured precession frequencies of the neutron to $^{199}$Hg atoms. The motional field of the mercury atoms induces a shift quadratic in $E$, $\delta_{\text{quad}}=-\gamma_{\text{Hg}}R^2E^2/(4c^2)$ \cite{Pignol2015-hn}. The mercury modulation frequency is shifted proportional to the intensity of the probe light \cite{Cohen-Tannoudji1962-ts}, which in turn is translated to a shift in the value of $\mathcal{R}$. Precessing polarized atoms of mercury produce a pseudomagnetic field that is felt by the cohabiting UCNs, $\bm{B}=-4\pi\hbar n_{\text{hg}} b_{\text{inc}} \bm{P}\sqrt{1/3}/(m_n\gamma_n)$ \cite{Abragam1982-dm}, where $b_{\text{inc}}=15.5~$fm \cite{Sears1992-xq} is the associated incoherent scattering length, $n_{\text{hg}}$ is the density of $^{199}$Hg atoms in the UCN precession chamber, and $\bm{P}$ is the polarization of the $^{199}$Hg atoms. Magnetic impurities and dirt may become lodged on the UCN precession chamber, and this also causes a shift in $\mathcal{R}$ and dilutes the precision of the measured neutron EDM. Finally, the EDM of the neutron itself was constrained to a statistical value of $d_n<1.1\times10^{-26}~$e$\cdot$cm \cite{Abel2020-jr}. These effects, in terms of the their contribution to the uncertainty on the neutron EDM, have been summarized in Table~\ref{tab10-1}.

\subsection{Crossing point analysis}
The shift in the value of $\mathcal{R}$, $\delta_{\text{grav}}$ given by Eq.~\ref{eq10-11a}, depends on the direction of the magnetic field. In the previous section, the effects that lead to shifts in the value of $\mathcal{R}$, were either constrained to small values (quoted in Table~\ref{tab10-1}) or compensated for directly, with the use of corrected parameters of $d_n^{\text{corr}}$ and $\mathcal{R}^{\text{corr}}$. The only effect that remains uncompensated for in the parameters of $d_n^{\text{corr}}$ and $\mathcal{R}^{\text{corr}}$ is from $\delta_{\text{grav}}$. This effect is compensated for by fitting the $d_n^{\text{corr}}$ as a linear function of $\mathcal{R}^{\text{corr}}$, whose slope is given by $\partial d_n^{\text{corr}}/\partial \mathcal{R}^{\text{corr}} = \hbar \gamma^2_{\text{Hg}}R^2B_0/(8\langle z \rangle c^2)$ \cite{Abel2020-jr}. Similarly, as evident from Eqs.~\ref{eq10-11} and \ref{eq10-11a}, $\mathcal{R}^{\text{corr}}$ can also be fitted as a linear function of $G_{\text{grav}}$. The sign of the slope is determined by the direction of the magnetic field. Such a linear fit is referred to as \emph{crossing point analysis}.

\begin{figure}[t]
\centering
\includegraphics[width=0.6\columnwidth]{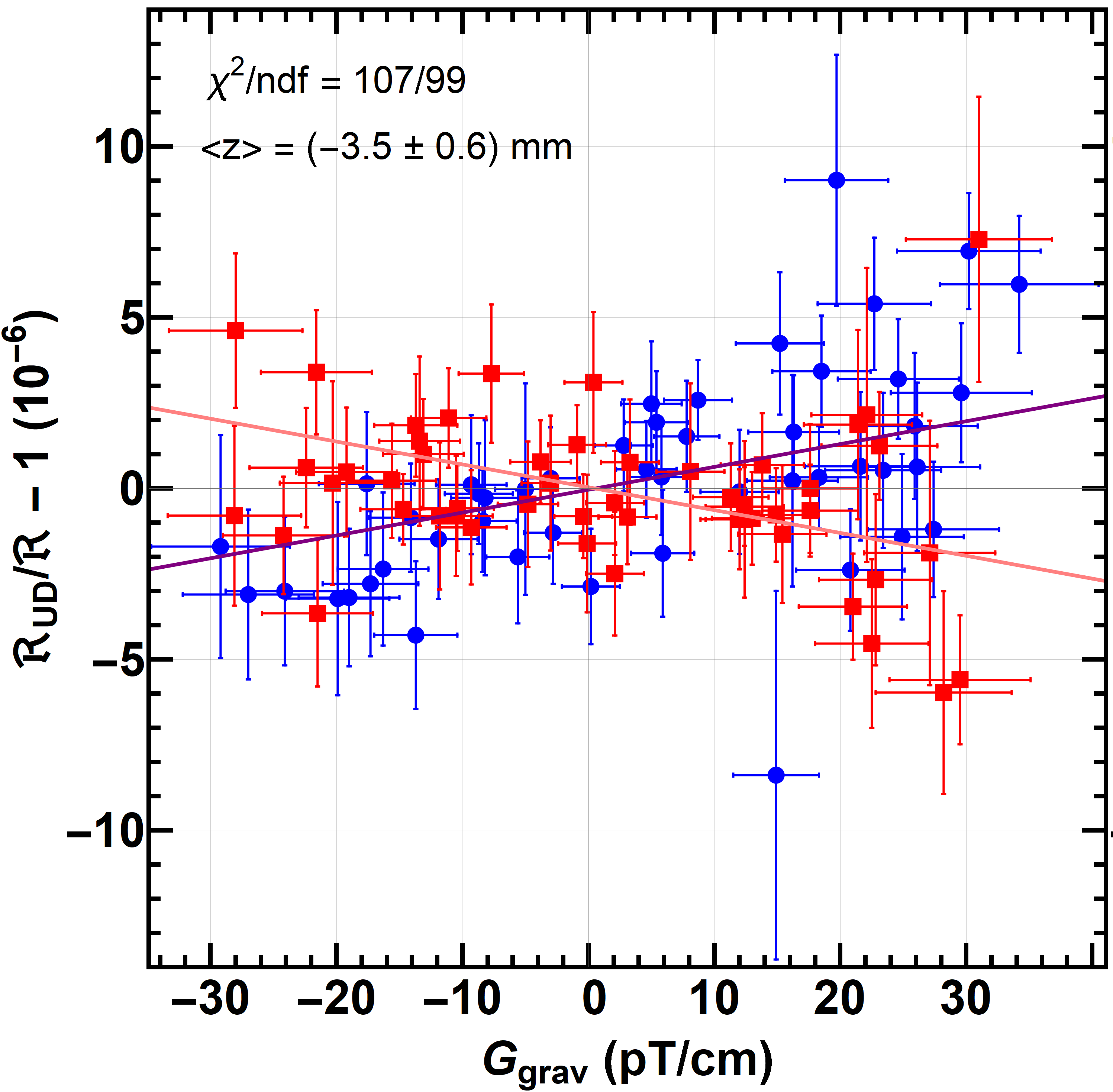}
\caption[]{\emph{Crossing point analysis}: The corrected value of $\mathcal{R}_{\text{UD}}$ has been plotted as function of the magnetic field gradient, $G_{\text{grav}}$. The red (square) data points correspond to those where $B_0$ direction was up, and the blue (circle) data points correspond to those where $B_0$ direction was down, and the central value of their corresponding linear fit is indicated by pink and purple lines, respectively.}
\label{fig10-0}
\end{figure}

\sloppy In this work, we have fitted, $\mathcal{R}_{\text{UD}}/\mathcal{R} - 1 = 2 E d^{\text{false}}_n /(\hbar \gamma_n B_0)$, as a linear function of $G_{\text{grav}}$, obtained from the $\{d^{\text{false}}_n,~\mathcal{R}^{\text{corr}}\}$ data in Fig.~4 of Ref.~\cite{Abel2020-jr}, jointly for the two directions of applied magnetic field. The subscript $UD$ here refers to the flipping of the direction of the electric field to obtain $d^{\text{false}}_n$ (not the to be confused with the arrow which usually refers to the direction of the magnetic field). The two fitted parameters involved are the center-of-mass offset parameter, $\langle z \rangle$, which dictated the slope, and a vertical-intercept, $\mathcal{R}_0$. The fit shown in Figure~\ref{fig10-0} yielded, $\chi^2/ndf=107/97$, and
\begin{eqnarray}
\langle z \rangle = (-3.5\pm0.6)~\text{mm}~;~\mathcal{R}_0 = (0.54\pm3.75)~\times10^{-6}.~\label{eq10-14}
\end{eqnarray}
Since the value of $\langle z \rangle$ also affects $G_{\text{grav}}$ via Eq.~\ref{eq10-11a}, we present our choice of $\langle z \rangle$ here that ultimately minimized $\chi^2$ associated with the crossing point fit. The value of $\langle z \rangle$ obtained here is consistent with the value of $\langle z \rangle_{\text{nEDM}} = (-3.9\pm0.3)~\text{mm}$ obtained in Ref.~\cite{Abel2020-jr}. The uncertainty associated with $\mathcal{R}_0$ is slightly larger than that associated with the value of $\mathcal{R}$ obtained in Ref.~\cite{Abel2020-jr} at the crossing point, due to the fact that the slope here is a free parameter. In order to accommodate for a larger uncertainty on the value of $\langle z \rangle$ obtained in this analysis compared to the nEDM analysis, the corresponding systematic effect in Table~\ref{tab10-1} has been scaled up accordingly.

In our fit, a system unperturbed by $n-n'$ oscillations, would yield an intersection point consistent with $\{0,0\}$. The presence of the $n-n'$ oscillation produces an additional slope, which is not correctable with a linear fit. The resultant intersection point upon the addition of $n-n'$ oscillation is thus displaced from the unperturbed scenario. After correcting for the $\delta_{\text{grav}}$ with the help of a crossing-point analysis, we can constrain such shifts by using the value of $\mathcal{R}^{\uparrow \downarrow}_{\text{UD}}$ so obtained where $G_{\text{grav}}=0$, where the arrow indicates the magnetic field direction. There could be additional offsets in $G_{\text{grav}}$, but we expect them to be negligible.

The shift due to $n-n'$ oscillation, according to Eq.~\ref{eq10-9}, depends on the relative difference between the precession frequency upon flipping the direction of the magnetic field. We can therefore use the values of $\mathcal{R}^{\uparrow \downarrow}_{\text{UD}}(G_{\text{grav}}=0)$ after being corrected for $\delta_{\text{grav}}$, for each of the magnetic field direction, yielding
\begin{eqnarray}
\frac{\delta \omega}{\omega} = \frac{\delta \mathcal{R}_{UD}}{\mathcal{R}} = \frac{\mathcal{R}^{\uparrow}_{\text{UD}} - \mathcal{R}^{\downarrow}_{\text{UD}}}{\mathcal{R}} ,~\label{eq10-15}
\end{eqnarray}
where $\mathcal{R}=3.8424574(30)$ \cite{Afach2014-ak}. Similarly, systematic uncertainties considered for the neutron EDM measurement in Table~\ref{tab10-1} also translate to an uncertainty on $\delta \mathcal{R}/\mathcal{R}$. The value of $\delta\omega/\omega$ thus obtained through Eq.~\ref{eq10-15} is
\begin{eqnarray}
\delta \omega/\omega = (-7\pm138_{\text{stat}}\pm27_{\text{sys}})\times10^{-8}.~\label{eq10-16}
\end{eqnarray}
Here, the statistical portion of the uncertainty arises purely from the crossing point fit, while the systematic portion is a aggregate of the systematic effects associated with $\delta \mathcal{R}/\mathcal{R}$, mentioned in Table~\ref{tab10-1}, combined in quadrature.

\begin{table}[t]
\caption[]{Sources of uncertainty contributions to the determination of $\delta\omega/\omega$ in Eq.~\ref{eq10-15} to be used in Eq.~\ref{eq10-9}. The top portion represents the statistical uncertainty, and the bottom portion indicates the quoted systematic uncertainties. \emph{Courtesy} Ref.~\cite{Abel2020-jr}.}
\label{tab10-1}
\centering
\begin{tabular}{l | c c}
\hline
\hline
{\small Errors from} & $\sigma_{d_n}$ & $\sigma_{\delta\mathcal{R}_{UD}/\mathcal{R}}$\\
& $\times(10^{-28})~\text{e}\cdot\text{cm}$ & $\times(10^{-8})$\\
\hline
Crossing point analysis & $107$ & $138$\\
\hline
$\langle z \rangle~(\times0.6~\text{mm}/0.3~\text{mm})$ & $14$ & $18$\\
$\hat{G}$ & $10$ & $13$\\
$\delta_{\text{T}}$ & $5$ & $6$\\
$^{199}$Hg-EDM & $0.1$ & $0.1$\\
Dipole contaminants & $4$ & $5$\\
Net rotational motion ($d^{\text{net}}$) & $2$ & $3$\\
$\delta_{\text{quad}}$ & $0.1$ & $0.1$\\
$G_{\text{grav}}$ drifts & $7.5$ & $10$\\
$\delta_{\text{light}}$ & $0.4$ & $0.5$\\
$\delta_{\text{inc}}$ & $7$ & $9$\\
\hline
TOTAL & $110$ & $140$\\
\hline
\hline
\end{tabular}
\end{table}

\section{Constraints on $n-n'$ oscillation}

\begin{figure}[t]
\centering
\includegraphics[width=\columnwidth]{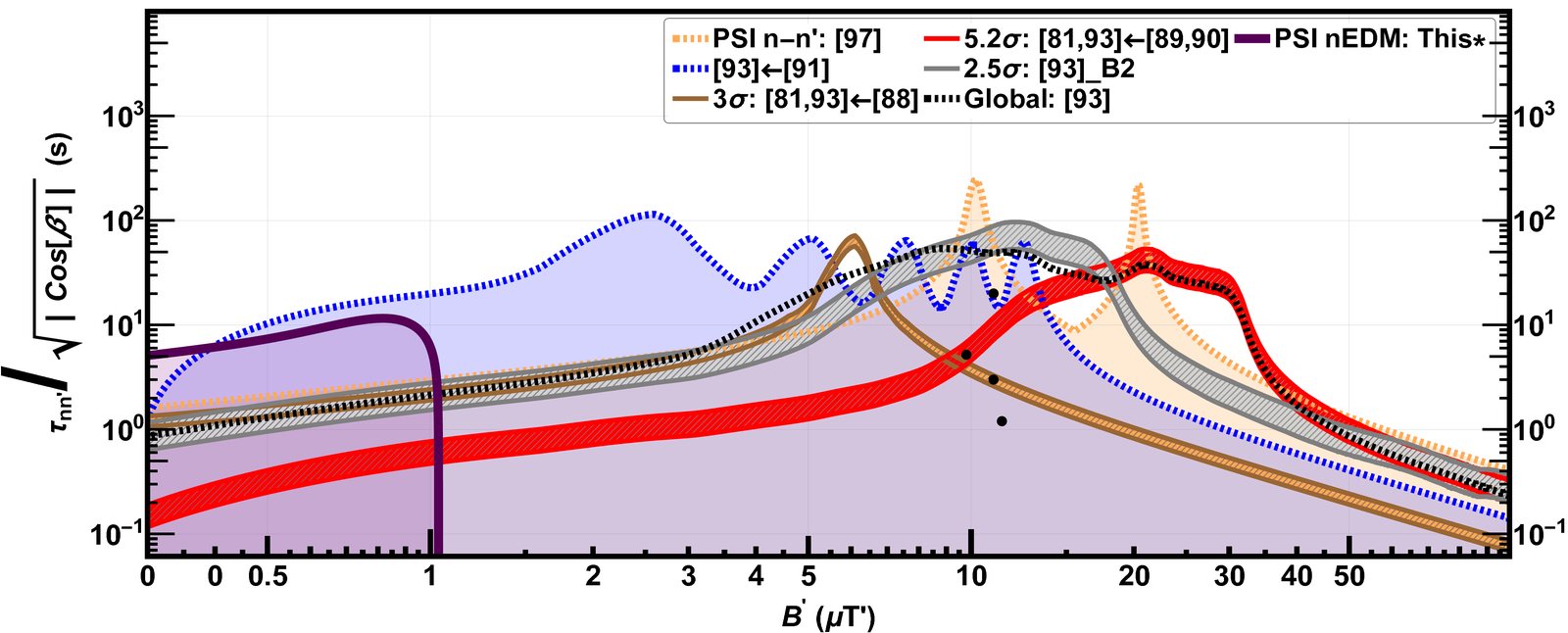}
\caption[]{Lower limits on the $n-n'$ oscillation time constant, $\tau_{nn'}/\sqrt{|\cos(\beta)|}$, at a 95\% C.L., while assuming $B'\ne0$. The lower limit from this effort has been shown as a solid purple curve. The dashed curves are previously imposed constraints: (a) the black curve is the global constraint calculated in Ref.~\cite{Berezhiani2018-df}, (b) the blue curve represents the constraint imposed using data in Ref.~\cite{Altarev2009-bv} by Ref.~\cite{Berezhiani2018-df}, and (c) the orange curve is the constraint imposed by Ref.~\cite{Abel2021-hp}. The three hatched regions are the anomalies (95\% C.L. contours): (i) the red region, is calculated in Refs.~\cite{Berezhiani2012-rq,Berezhiani2018-df} from the $5.2\sigma$ anomaly in Refs.~\cite{Serebrov2008-jg,Serebrov2009-yf}; (ii) the brown region is calculated in Refs.~\cite{Berezhiani2012-rq,Berezhiani2018-df} from the $3\sigma$ anomaly in Ref.~\cite{Ban2007-dt}; and (iii) the gray region comes from the $2.5\sigma$ anomaly observed in the B2 series of Ref.~\cite{Berezhiani2018-df}. The black dots indicate the solution consistent with the statistically significant signals as reported in Ref.~\cite{Berezhiani2012-rq}.}
\label{fig10-1}
\end{figure}

Neutron to mirror-neutron oscillation changes the slope of $\mathcal{R}^{\uparrow\downarrow}_{\text{UD}}$ as a function of $G_{\text{grav}}$ such that a simple linear fit to the data cannot be used. The null hypothesis condition involving no $n-n'$ oscillation leads to the value of $\mathcal{R}^{\uparrow\downarrow}_{\text{UD}}(G_{\text{grav}}=0)=0$ with a very good $\chi^2/ndf$. The measurement of $\delta\omega/\omega$ in Eq.~\ref{eq10-16} is statistically consistent with a null hypothesis. Therefore, no $n-n'$ oscillation was observed but we can constrain the process. In order to obtain a constraint on the oscillation time constant, $\tau_{nn'}$, using Eq.~\ref{eq10-9}, a constraint on the value of $\sqrt{1/(\delta\omega/\omega)} = \tau_{nn'}/(\sqrt{|\cos(\beta)|} \sqrt{f_{d}(\eta)})$ was calculated numerically at a $95\%$ C.L. to be,
\begin{eqnarray}
\frac{\tau_{nn'}}{\sqrt{|\cos(\beta)|} \sqrt{f_{d}(\eta)}} > 609.~\label{eq10-15a}
\end{eqnarray}
A constraint upon $\tau_{nn'}/\sqrt{|\cos(\beta)|}$ was obtained by scaling this value with the function $\sqrt{f_{d}(\eta)}$ defined in Eq.~\ref{eq10-9}. The resulting parameter space is defined by $\tau_{nn'}/\sqrt{|\cos(\beta)|}$ and $B'$, given that the applied magnetic field was fixed to $\langle B_0 \rangle\approx\pm\SI{1.036}{~\micro T}$. This new constraint has been plotted in Figure~\ref{fig10-1} along with constraints and anomalies from previous efforts.

The functional form of the scaling function, $f_d(\eta)$, in Eq.~\ref{eq10-9}, implies that the values of $\tau_{nn'}$ are real under the condition $B'<\langle B_0 \rangle$ or $B'<-\langle B_0 \rangle$. Therefore, these constraints are valid in the range of $B'<\SI{1.036}{~\micro T'}$. The singularity at $B'=\langle B_0 \rangle$ in Eq.~\ref{eq10-9} was eliminated by using the approximation in Eq.~\ref{eq10-9b}. This approximation used a value for mean free flight time, $\langle t_f \rangle$, which depends on the free spin precession time. UCNs were allowed to freely precess in Ref.~\cite{Abel2020-jr} for a time period of $t^*_s=180~$s. The values of $\langle t_f \rangle_{\tiny (t^*_s=180~\text{s})}=0.0628(27)~$s and $\langle t^2_f \rangle_{\tiny(t^*_s=180~\text{s})}=0.00653(69)~$s$^2$ from Refs.~\cite{Abel2021-hp,Mohanmurthy2019-ju} were used here, since the data employed here was collected in the same apparatus.

Furthermore, the range of $B'$ for which the constraints have been plotted in Figure~\ref{fig10-1}, $B'\in(0.36,100)\SI{}{~\micro T}$, are the same as in Ref.~\cite{Abel2021-hp}. The lower limit of the region of interest comes from the condition that $\omega'\langle t_f \rangle \gg 1$ which gives $B'>\SI{0.36}{~\micro T'}$ \cite{Abel2021-hp}, while the upper limit of the region of interest, $B'\lesssim\SI{100}{~\micro T}$ comes from the constraints on UCN losses in the Earth's magnetic field \cite{Berezhiani2009-hb,Berezhiani2018-df,Abel2021-hp}.

\section{Discussion of the Results}

For the first time, a constraint on $n-n'$ oscillation was obtained by using the result of an experiment searching for the neutron electric dipole moment. In order to compensate for the systematic effects arising from the magnetic field gradients, data was collected by Ref.~\cite{Abel2020-jr} using both orientations of the magnetic field, while holding its magnitude nearly constant. This allowed us to use the neutron EDM result to extract the difference between the neutron precession frequency when the magnetic field is flipped.

The constraint from this work shown in Figure~\ref{fig10-1} can be interpreted, at 95\% C.L., as
\begin{eqnarray}
\frac{\tau_{nn'}}{\sqrt{|\cos(\beta)|}} > \SI{5.7}{~s},~\SI{0.36}{~\micro T'}<B'<\SI{1.01}{~\micro T'}.\label{eq10-16a}
\end{eqnarray}
The constraint from this effort has been plotted along with previous results that also constrained the parameter of $\tau_{nn'}/\sqrt{|\cos(\beta)|}$ in Figure~\ref{fig10-1}. Other similar parameters used the asymmetry between the relative number of neutrons that survived storage upon flipping the direction of the applied magnetic field.

In this effort, we are concerned with terrestrial sources of $B'$ that are fixed to the rotating frame of the Earth. Thus, the angle between the applied $\bm{B_0}$ and the ambient $\bm{B'}$, $\beta$, is assumed to be a constant over time. Furthermore, the magnitude of the ambient mirror magnetic field, $|\bm{B'}|$, was also assumed to be a constant over time. However, studying the sidereal variations of the difference in precession frequencies ($\partial \omega$ in Eq.~\ref{eq10-15}) could also be a viable means to search for $n-n'$ oscillation in the case of galactic origins of the ambient mirror magnetic field.

All but one experiment thus far, were performed at the Institute Laue-Langevin (ILL) \cite{Ban2007-dt,Serebrov2008-jg,Altarev2009-bv,Serebrov2009-yf,Berezhiani2012-rq,Berezhiani2018-df}, while the most recent effort was carried out at PSI \cite{Abel2021-hp}. This effort uses data from Ref.~\cite{Abel2020-jr} which was also collected at PSI. We here present the constraints from all the efforts on a single plot in Figure~\ref{fig10-1}, despite the possibility of ambient mirror magnetic fields being different at the two locations. However, under the natural assumption of approximate rotational symmetry of a mirror magnetic field bound to the Earth \cite{Berezhiani2009-hb}, the components of $\bm{B'}$ would vary negligibly between the two locations \cite{Abel2021-hp}.

Due to the fact that the PSI nEDM effort used a magnetic field of $B_0\approx\SI{1.036}{~\micro T}$, which is much lower than previous efforts that searched for $n-n'$ oscillation, this constraint is the best available for low values of the mirror magnetic field, in the range of $\SI{0.36}{~\micro T'} < B' < \SI{0.40}{~\micro T'}$. However, the best local constraint in the same parameter space is $(\tau_{nn'}/\sqrt{\cos(\beta)}) > 241~$s at $B'=\SI{10.2}{~\micro T'}$ (95\% C.L.) from Ref.~\cite{Abel2021-hp}, shown as an orange dashed line in Figure~\ref{fig10-1} and the best overall constraint is $(\tau_{nn'}/\sqrt{\cos(\beta)})>27~\text{s},~\SI{6}{~\micro T'}<B'<\SI{25}{~\micro T'}$ (95\% C.L.) from Ref.~\cite{Berezhiani2018-df}, shown as a black dashed line in Figure~\ref{fig10-1}.

This effort excludes portions of all three anomalies, indicated in Figure~\ref{fig10-1}, in the range $B'<\SI{1.0}{~\micro T'}$. But, these regions have been previously excluded by Ref.~\cite{Altarev2009-bv}. There are regions, in the range of $\SI{4}{~\micro T'}<B'<\SI{37}{~\micro T'}$, where at least two of these anomalies overlap, that are still not excluded. However, reiterating Ref.~\cite{Abel2021-hp}, the three anomalies are not mutually consistent, \emph{i.e.} considering any of the two anomalies together excludes the third. Future planned neutron EDM experiments \cite{Piegsa2013-jl,Serebrov2017-ey,Wurm2019-ys,Kuchler2019-fb,Ito2018-zm,Ahmed2019-oo,Ayres2021-uk} may use larger magnetic fields and achieve better sensitivity to the neutron EDM. That will allow these proposed experiments to be sensitive to a larger region of the parameter space in Figure~\ref{fig10-1}.


\funding{One of the authors, P. M., would like to acknowledge support from Sigma Xi grants \# G2017100190747806, \# G2019100190747806, and the Ivy Plus exchange program. J. W. is supported by DOE grant \# DE-SC00-14448. A. R. Y. is supported by NSF grants \# 1615153, \# 1914133 and DOE grant \# DE-FG02-ER41042.}

\acknowledgments{We would like to acknowledge the office resources provided by the James Frank Institute and the Mansueto Library at the University of Chicago.}

\end{paracol}
\reftitle{References}

\end{document}